\documentstyle[11pt,paspconf,psfig,epsf]{article}
                                                 
\def \hi {H\,{\sc i~}}

\def \kmsws {$\rm km\,s^{-1}$}
\def \kmswss {$\rm km\,s^{-1}\;$}

\begin{document}
                
\title{The gaseous Milky Way halo}
                                  
\author{Peter M.W. Kalberla, J\"urgen Kerp}
\affil{Radioastronomisches Institut der Universit\"at Bonn, Auf dem
H\"ugel 71, D-53121 Bonn, Germany}
                                  
\begin{abstract}
In 1998 several papers claim the detection of an ubiquitous gaseous phase
within the Galactic halo.
Here we like to focus on the detections of X-ray emitting gas within the
Galactic halo as well as the discovery of a pervasive neutral Galactic halo
gas. We discuss critically the major differences between 
the recent publications as well as the limitations of the analyses. 

\end{abstract}
              
                                       
\section{The main constituents of the ISM }
                      
The detection of neutral interstellar clouds at large $z$-distances
(M\"unch 1957) led to the hypothesis of a hot
gaseous Galactic halo (Spitzer 1956).
This high temperature ($T\simeq 10^6$ K) low volume density gas was proposed
to confine the neutral clouds high above the Galactic disk.
Alternatively, a low temperature, almost neutral halo
($T\simeq 10^4$ K) was postulated by Pikelner \& Shklovsky (1958). Their model
predicted emission lines of neutral species with
velocity dispersions of 70\,\kmsws due to turbulent motions.

In the early
fifties optical polarization studies revealed that the Galaxy hosts
an interstellar magnetic field, with a field strength of a few $\mu$G.
This is, on the average, oriented parallel to the Galactic disk.
Several years later, radio continuum surveys (Beuermann et al. 1985) gave clear evidence for 
radio
synchrotron emission originating at distances of a few kpc above the Galactic plane.
This indicates that magnetic fields are also present within the Galactic halo.
Already in 1966 Parker pointed out that magnetic fields must be associated 
with gaseous counterparts.
                                       
UV absorption line measurement with the {\it Copernicus\/} satellite showed the presence
of highly-ionized species within the Galactic halo.
However, the detection of an ubiquitous gaseous component was not easy to establish.
It took nearly 40 years, since Spitzer's prediction.
The {\it ROSAT\/} all-sky survey revealed the presence of an ubiquitous hot
X-ray emitting plasma within the Galactic halo.
Pietz et al. (1998) and Kerp et al. (1999) showed that the X-ray emitting halo gas
is smoothly distributed across the whole sky.
They performed a quantitative correlation the new Leiden/Dwingeloo HI 21-cm line survey and the 
{\it ROSAT\/} all-sky survey.
These analyses gave evidence that the Galactic halo is the brightest diffuse soft X-ray source.

Also recently, evidence for
an extended neutral Galactic halo was presented by Albert et al. (1994).
Lockman \&  Gehman (1991) found \hi gas with a velocity
dispersion of up to 35\,\kmswss towards the Galactic poles, which was
interpreted as emission originating from a Galactic halo.
Kalberla et al. (1998a) disclosed the existence of a pervasive \hi component with a velocity
dispersion of 60 \kmswss from the Leiden/Dwingeloo Survey 
(LDS, Hartmann \& Burton 1997).

The X-ray as well as the neutral gas within the Galactic halo are consistent with a 
hydrostatic equilibrium model of the Milky Way on large 
scales (Kalberla \& Kerp 1998).
The X-ray data constrain the extent of the gaseous halo, while the high-velocity dispersion
component determines the pressure balance of the equilibrium model.
According to this equilibrium model of Kalberla \& Kerp (1998), the Galactic halo has a vertical 
scale
height of 4.4\,kpc and a radial scale length of 15\,kpc.
These values are in agreement with those derived from the distribution of the highly-ionized 
atoms
within the Galactic halo published by Savage et al. (1997): $h_z(\rm{Si}\sc {iv}) = 5.1 
(\pm0.7)$\,kpc,
$h_z({\rm C\sc {iv}}) = 4.4 (\pm 0.6) $\,kpc and $h_z({\rm N \sc{v}}) = 3.9 (\pm 1.4)$\,kpc.
Kalberla et al. (1998a) as well as Savage et al. (1997) find that the observational data are 
fitted best by 
assuming a co-rotation of the Galactic disk and halo in addition to a turbulent motion of the 
gas with an
averaged velocity dispersion of 60\,\kmswss. 
Taking all in one, it is possible to construct a consistent model of the Galactic halo,
consisting of gas, magnetic fields and cosmic rays.

In this paper we discuss the evidence for a gaseous halo in some detail.
In Sect. 2 we describe observational evidence for the existence of soft X-ray plasma within the 
Galactic halo.
In Sect. 3 we focus on \hi observations and model calculations of the Galactic disk/halo.
In Sect. 4 we discuss the stability of a multi-phase disk-halo system.
                     
\section{X-ray emission from the halo }
                                       
Since the first detection of the diffuse soft X-ray  background in 1968
(Bowyer, Field and Mack) it was a matter of debate whether the radiation
originates within the Galactic halo or in the local vicinity of the Sun.
Because of the strong energy dependence of the photo-electric cross section on
energy ($\sigma\,\propto\,E^{-\frac{8}{3}}$) the soft X-ray emission is
strongly attenuated by the ISM on the line of sight.
In the so called 1/4\,keV band the cross section is about 
$\sigma\,=\,0.5\,10^{-20}\,{\rm cm^2}$; thus, towards most of the 
high latitude sky we have
at least an opacity of unity.

With the {\it ROSAT\/} X-ray telescope it was possible
to study the diffuse soft X-ray background emission on a high signal-to-noise
ratio. The angular resolution of 30" was sufficient to reveal individual 
shadows of clouds in front of an X-ray emitting source.

Here, we like to compile the major findings of two recent papers on the Galactic X-ray halo.
Both are based on the {\it ROSAT\/} all-sky survey data, both assume that the soft
X-ray background radiation is a superposition of the local hot bubble,
the Galactic halo and extragalactic background emission.
Despite these similarities, they derive a totally different Galactic halo X-ray intensity 
distribution.
What may cause these differences and what are the implications?

{\it The data-set:\/} Snowden et al. (1998) analyzed {\it ROSAT\/} 1/4\,keV band data,
while Pietz et al. (1998) studied the X-ray intensity distribution of the {\it ROSAT\/} 3/4\,keV 
and 
1/4\,keV energy band. 
Because of the strong energy dependence of the absorption cross section,
the 3/4\,keV emission is only attenuated weakly, while the 1/4\,keV 
photons are strongly absorbed by 
the same amount of interstellar matter.
For example a neutral hydrogen column density of $N_{\rm \hi}\,=\,1.5\,10^{20}\,{\rm cm^{-2}}$ 
attenuates the 3/4\,keV radiation by only 15\%, while the same column density absorbs 75\% of 
the 1/4\,keV 
photons.
This major difference indicates, that the 3/4\,keV is more appropriate 
to study the intensity distribution of a hot Galactic halo gas.

{\it The plasma temperature:\/}
Because of the moderate energy resolution of the {\it ROSAT\/} PSPC detector, the practical way 
to
determine the plasma temperature across the Galactic sky is to evaluate energy-band ratios.
Especially the 1/4\,keV to 3/4\,keV band ratio is very sensitive to variations of the plasma
temperature. A ${\rm log}(T{\rm [K]})\,=\,6.0$ plasma will emit most of the photons in the 
1/4\,keV 
range, but a minor increase in temperature, for instance to ${\rm log}(T{\rm [K]})\,=\,6.3$
produces about an order of magnitude more 3/4\,keV photons than the ${\rm log}(T{\rm 
[K]})\,=\,6.0$ 
plasma.

Snowden et al. (1998) derived a LHB temperature of ${\rm log}(T{\rm [K]})\,=\,6.07$ while the
Galactic halo temperature is lower and about ${\rm log}(T{\rm [K]})\,=\,6.02$.
Both plasma do {\em not} contribute significantly to the 3/4\,keV energy range.
In contrast to this finding, Pietz et al. (1998) claimed that the temperature of
the LHB is about ${\rm log}(T{\rm [K]})\,=\,5.85$ and that of the halo
${\rm log}(T{\rm [K]})\,=\,6.2$. 
The LHB plasma does not emit significant amounts of 3/4\,keV
photons, while the halo plasma accounts for about half of the observed 3/4\,keV emission.
Accordingly, the gap between the observed 3/4\,keV intensity and the known sources of diffuse 
3/4\,keV
radiation, already identified within the Wisconsin survey 
(McCammon \& Sanders 1990), is overcome by the approach of Pietz et al.

{\it The method:\/} Snowden et al. fitted scatter-diagrams
(\hi column density $N_{\rm \hi}$ versus X-ray intensity $I_{\rm X-ray}$). 
For this approach it is necessary to assume an {\it a priori\/} plasma temperature to determine 
the
photo-electric cross section of the ISM.
Snowden et al. analyzed the {\it ROSAT} 1/4\,keV radiation splitted into the so-called R1 and R2
energy bands.
In the second step, they derived the temperature of the plasma components, by evaluating the 
R1/R2
energy-band ratio.
The R1/R2 band ratio is only a weak function of temperature. Using the same plasma temperatures 
as
in the example above we find for ${\rm log}(T{\rm [K]})\,=\,6.0$ 
a ratio R1/R2 = 0.9, while for
${\rm log}(T{\rm [K]})\,=\,6.3$,  R1/R2 = 0.6.
The uncertainties within of the {\it ROSAT} R1 and R2 band are about 10\%, accordingly the 
R1/R2-band
ratio has a statistical uncertainty of about 20\%.
Thus, it is a difficult task to constrain the plasma temperature by analyzing
the R1/R2-band ratio.
In the final step, they derived
the Galactic halo intensity by inverting the radiation transfer equation.
Within the 1/4\,keV band the product of $N_{\rm \hi}$ and $\sigma$ is close to or larger as
unity, the quotient $I_{\rm halo}\,=\frac{\,I_{\rm LHB}\,-\,I_{\rm ex}\,
e^{-\sigma\,N_{\rm \hi}}}{e^{-\sigma\,N_{\rm \hi}}}$ becomes uncertain, because the divisor is
a small value.

The approach of Pietz et al. was totally different.
They found evidence for a large scale 3/4\,keV intensity gradient 
between the Galactic center
and the Galactic anti-center direction.
Because the 3/4\,keV radiation is only to a minor fraction attenuated by 
the Galactic 
interstellar matter towards
the high latitude sky, they attributed this gradient to a variation of the
emissivity of the Galactic halo plasma.
They modeled this 3/4\,keV gradient and optimized the shape of the Galactic halo to fit the 
3/4\,keV
diffuse X-ray background intensity distribution.
The attenuation of the 3/4\,keV radiation is much weaker than in the 1/4\,keV energy range, but
present. Pietz et al. estimated the Galactic halo plasma temperature by the spectral fit of the
diffuse X-ray background data of a deep pointed PSPC observation.
This is the main source of uncertainty, because it is unknown whether this pointed observation
is towards a representative direction or not.
To overcome this source of uncertainty, they investigated the 1/4\,keV to 3/4\,keV energy ratio, 
which
is, as shown above, a sensitive measure of the plasma temperature. 
They cross-correlated the 1/4\,keV Galactic halo intensities of 
Kerp et al. (1999) with that of the
modeled 3/4\,keV sky and deduced a plasma temperature of ${\rm log}(T{\rm 
[K]})\,=\,6.2\,\pm\,0.02$
(compare their Fig.\,10).
In following step, they scaled the emissivity of 3/4\,keV model to the 1/4\,keV energy range.

{\it The r\^ole of the extragalactic background radiation:\/}
The radiation transfer equation of Snowden et al. and Pietz et al. is the same.
An important, but up to now not discussed diffuse soft X-ray 
component is the extragalactic 
background.
The 1/4\,keV to 3/4\,keV energy-band ratio is not only a sensitive measure of the plasma 
temperature,
but also a measure for the brightness of the extragalactic background.
As pointed out by Kerp \& Pietz (1998), if the extragalactic background spectrum is steep
$I_{\rm extra}(E)\,\simeq\,E^{-2}$ (Hasinger et al. 1993) and the intensity within the 1/4\,keV 
band
high $I_{\rm extra}\,\simeq\,4.4\,10^{-4}\,{\rm cts\,s^{-1}\,arcmin^{-2}}$ 
(Cui et al. 1996),
the quantitative analysis of the {\it ROSAT\/} energy-band ratios 
allows only a low Galactic halo
plasma temperature (${\rm log}(T{\rm [K]})\,<\,6.1$).
If one assumes a flatter extragalactic background spectrum 
($I_{\rm extra}(E)\,\propto\,E^{-1.5}$, Gendreau et al. 1995) and a fainter 1/4\,keV intensity 
($I_{\rm extra}\,\simeq\,2.3\,10^{-4}\,
{\rm cts\,s^{-1}\,arcmin^{-2}}$, Barber et al. 1996), a higher Galactic halo plasma temperature 
is consistent with the energy band ratios.

Snowden et al. adopt the steep spectrum high 1/4\,keV intensity set of parameters, and found 
accordingly a low temperature halo. 
Moreover, this choice of parameters for the extragalactic background determines that 66\% of 
the 
3/4\,keV diffuse X-ray emission is of extragalactic origin, 
while low plasma temperatures derived for the 
Galactic halo and the LHB {\em cannot} account for the rest of the 3/4\,keV emission.
Thus, a significant fraction of the 3/4\,keV diffuse X-ray radiation is not investigated.

Pietz et al. used the flatter background spectrum 
and low intensities for the extragalactic background, 
the second set of parameters, 
which allows a higher temperature for the Galactic
halo plasma.
The difference between the 3/4\,keV extragalactic background level (also 66\% of the total 
observed
3/4\,keV intensity) and the observed 3/4\,keV intensity is attributed 
to the Galactic halo plasma.
Apparently, the 3/4\,keV gap is filled with the Galactic halo plasma emission.

{\it The model predictions:\/} Snowden et al. (1998) confirm the LHB model of 
(Snowden et al. 1990) only slightly modified because of the Galactic halo emission.
In the Snowden et al. (1998) model the Galactic halo 
contains some patches of soft X-ray emitting halo gas,
which are distributed across the high latitude sky.
The 3/4\,keV emission is not entirely explained by the model, because the gap between the
extragalactic X-ray background intensity level and the observed intensity is still present.

The Pietz et al. model explains entirely the 3/4\,keV and 1/4\,keV X-ray background emission 
towards
high latitudes.
The derived Galactic halo plasma temperature of ${\rm log}(T{\rm [K]})\,=\,6.2$ emits most of 
the
soft X-ray photons in the 1/4\,keV energy band.
Accordingly, they scaled the emissivity of the 3/4\,keV model to that of the 1/4\,keV X-ray 
emission.
They attribute the difference between the large-scale averaged observed 1/4\,keV and modeled
1/4\,keV Galactic halo emission to the LHB.
The derived LHB emission is smoothly distributed across the sky.

{\it Conclusions:\/} Snowden et al. confirmed the LHB model, in which a displacement of X-ray
plasma and $N_{\rm HI}$ produces the apparent Galactic plane to Galactic pole X-ray intensity
gradient.
They found evidence for some low temperature soft X-ray emitting plasma patches within the 
Galactic 
halo gas.
They conclude that temperature and intensity of the halo emission must 
be highly variable which excludes any pervasive Galactic halo plasma.

Pietz et al. constructed a model of the X-ray halo which is consistent solution for the detected
diffuse soft X-ray emission across the entire {\it ROSAT\/} X-ray energy range.
They proposed, that the 3/4\,keV emission is a superposition of the Galactic halo and 
extragalactic 
background emission.
The Galactic halo plasma appears to be isothermal on large angular scales, accordingly it is 
possible 
to scale the 3/4\,keV model to the 1/4\,keV energy regime.
They derived a smooth intensity distribution of the LHB, which 
is significantly cooler than the
Galactic halo plasma.
The smooth LHB intensity distribution is a function of the Galactic latitude.
Towards the northern Galactic pole, the LHB is about 1.5 times brighter than towards the 
southern Galactic pole.
The residuals between the observed and modeled X-ray intensity distributions 
show some individual
excess X-ray emitting structures which are identified as star forming regions, 
supernova remnants or
attributed to the high-velocity cloud phenomenon (Kerp et al. 1999).

The Pietz et al. Galactic X-ray halo model predicts the existence of a hot gaseous phase within 
the Galactic halo.
This hot gas phase is the environment of cosmic-rays, magnetic fields and neutral clouds, 
accordingly,
we can prove the X-ray model parameters by the quantitative comparison with recent all-sky 
surveys
of \hi, radio continuum and $\gamma$-ray emission (Kalberla \& Kerp, 1998).

\section{\hi gas in the halo }
                                      
On large scales, the vertical distribution of \hi gas in the Galaxy can be 
characterized by a layered structure (Dickey \& Lockman 1990). 
The layers which are associated with the disk have an exponential scale 
height of up to 400 pc. Until recently there was also agreement that 
\hi gas associated with the halo must have a scale height of $ 1 < h_z < 
1.5 $ kpc (e.g. Kulkarni \& Fich 1985 or Lockman \& Gehman 1991). Such a 
scale height corresponds to a component with a velocity dispersion of 
35 \kmsws. This was questioned by 
Westphalen (1997) and Kalberla et al. (1998a) who found a velocity 
dispersion of 60 \kmsws, corresponding to a scale height of 4.4 kpc. 
Here we discuss the major differences between both approaches. 

{\it The data-set:\/} 
A component with a velocity dispersion of $\sigma \approx $ 35 \kmsws 
was derived by different authors 
from \hi profiles extracted from the Bell Laboratories \hi survey 
(BLS, Stark et al. 1992). The main reason for using this horn-antenna 
for such a kind of analysis was the high main beam efficiency of 
92\%, which means that the observations are only little affected 
by stray radiation from the antenna side-lobes. 

A high velocity dispersion component  of $\sigma \approx $ 60 \kmsws was 
found by Westphalen (1997) using the LDS. This 
survey was observed with a parabolic dish. Usually, 
such a telescope suffers from spurious emission which disables any 
analysis of \hi components with high-velocity dispersion. 
In the case of the LDS, however, the stray radiation was removed in a numerical way
by Hartmann et al. (1996). These authors claim that they 
reduce spurious emission on average by two orders of magnitude. 
They demonstrated, that even reflections from the ground 
can affect \hi observations. Subsequently, such effects have been 
corrected by Kalberla et al. (1998a) who suggested that systematic 
uncertainties due to residual stray radiation in wings of the \hi spectra
are below 15 mK.

{\it Data reduction:\/} 
\hi emission lines with high-velocity dispersion may be seriously affected by 
the instrumental baseline and the correction algorithm which was used.
Individual BLS profiles were corrected by fitting a second order 
polynomial to emission free profiles regions which have been determined 
in an iterative way. In case of the LDS, a third 
order polynomial fit was applied after determining the emission free 
regions with polynomials up to $5^{th}$ order.

%
\begin{figure}[t]
\centerline{\hspace{-1.5cm}}
\psfig{figure=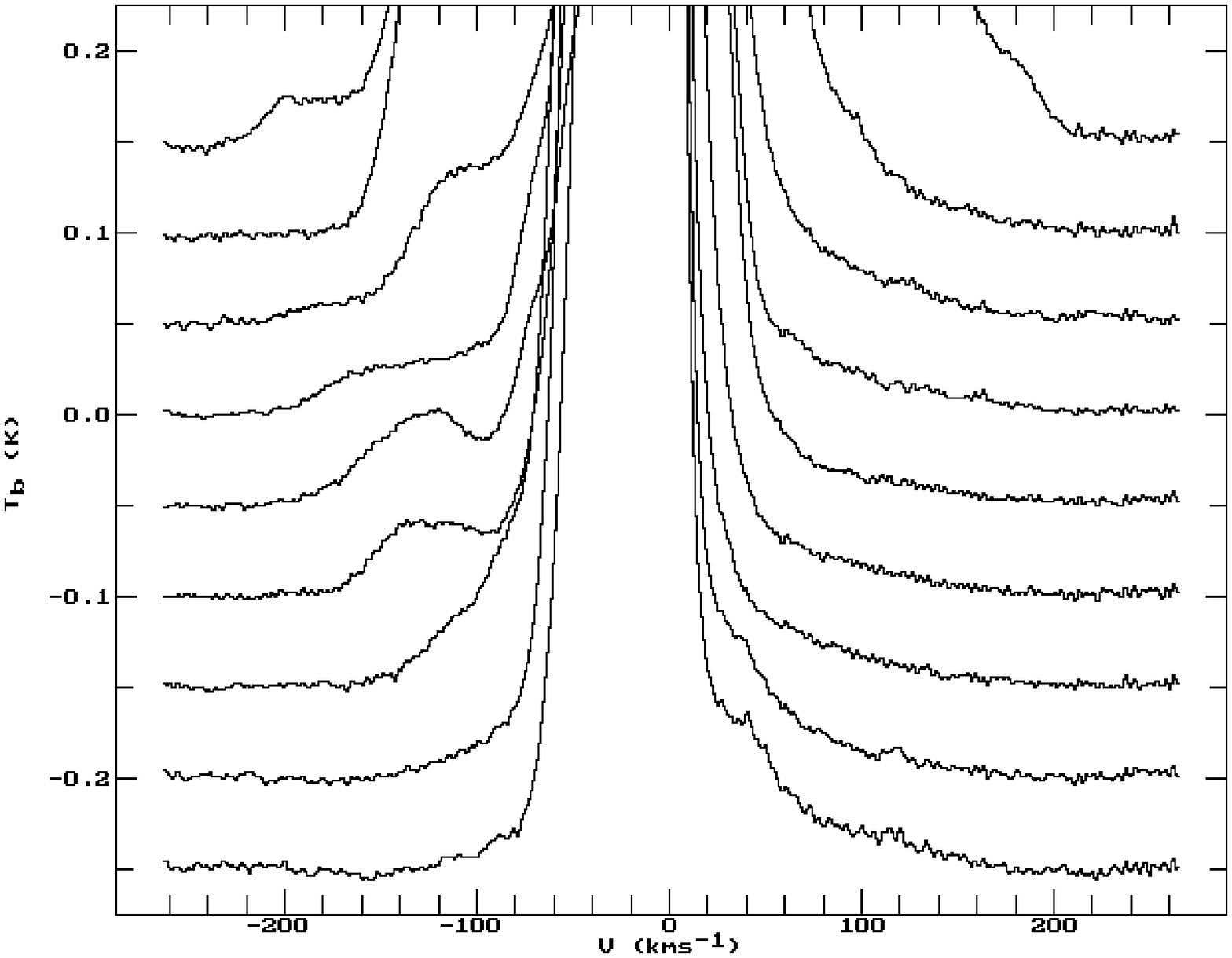,width=13.2cm,angle=-90
,bbllx=50pt,bblly=100pt,bburx=530pt,bbury=730pt}
\vspace*{-10.4875cm}
\centerline{\hspace*{-1.5cm}}
\psfig{figure=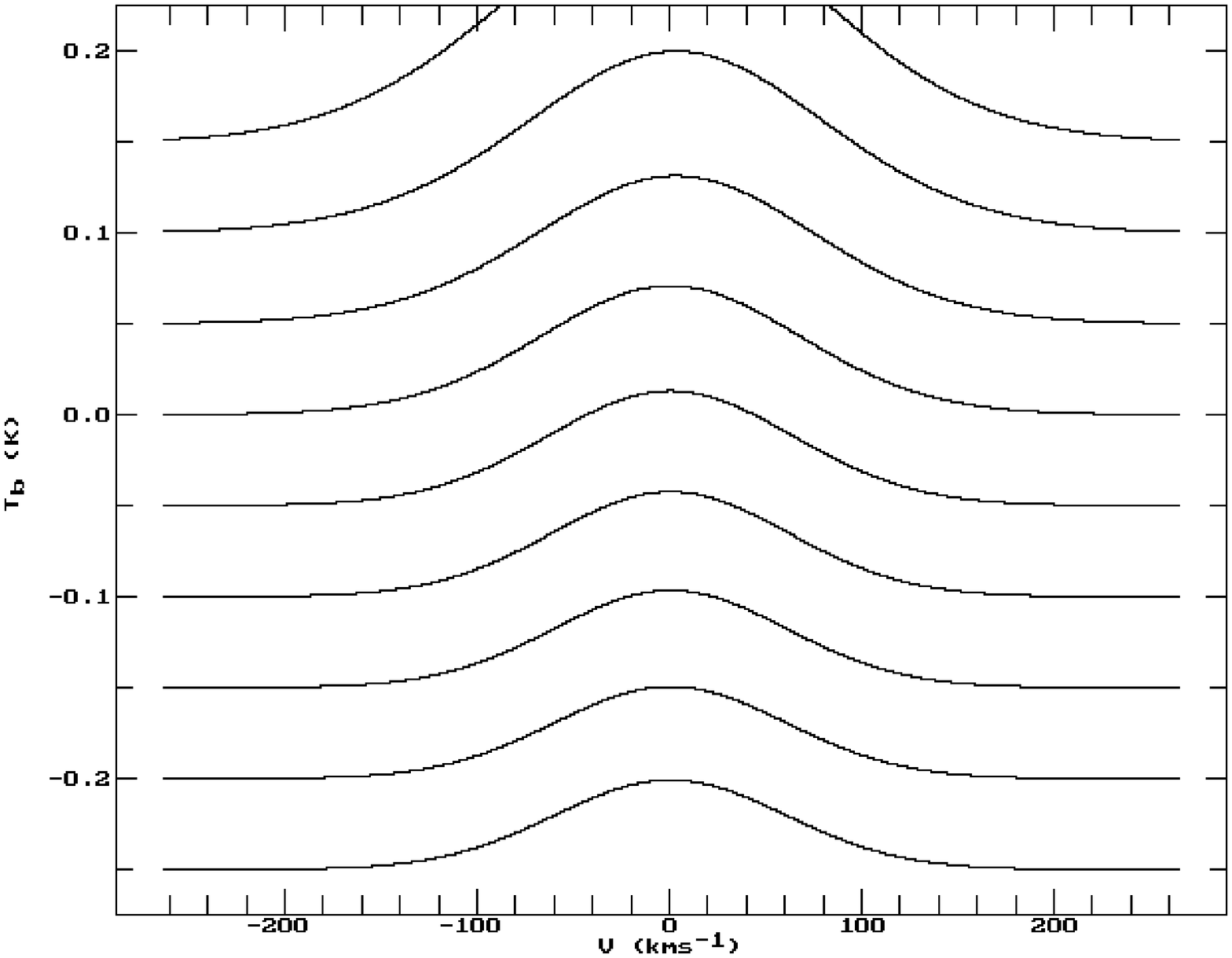,width=13.2cm,angle=-90
,bbllx=50pt,bblly=100pt,bburx=530pt,bbury=730pt}
\caption[]{
\hi 21-cm line spectra from the Leiden/Dwingeloo Survey (LDS)
averaged over all longitudes and in latitude across $10\deg$.
The bottom profile is centered at $b=85\deg$,
the top one
at $b=5\deg$ in steps of $10\deg$. The zero levels of subsequent
profiles are separated by 50 mK.
The solid lines mark 
the \hi emission of a co-rotating halo with a vertical
scale height of $h_z=4.4$ kpc and a radial scale length of
$A_1=15$ kpc.
\label{fig1} }
\end{figure}

%
\begin{figure}[h]
\centerline{\hspace{-1.5cm}}
\psfig{figure=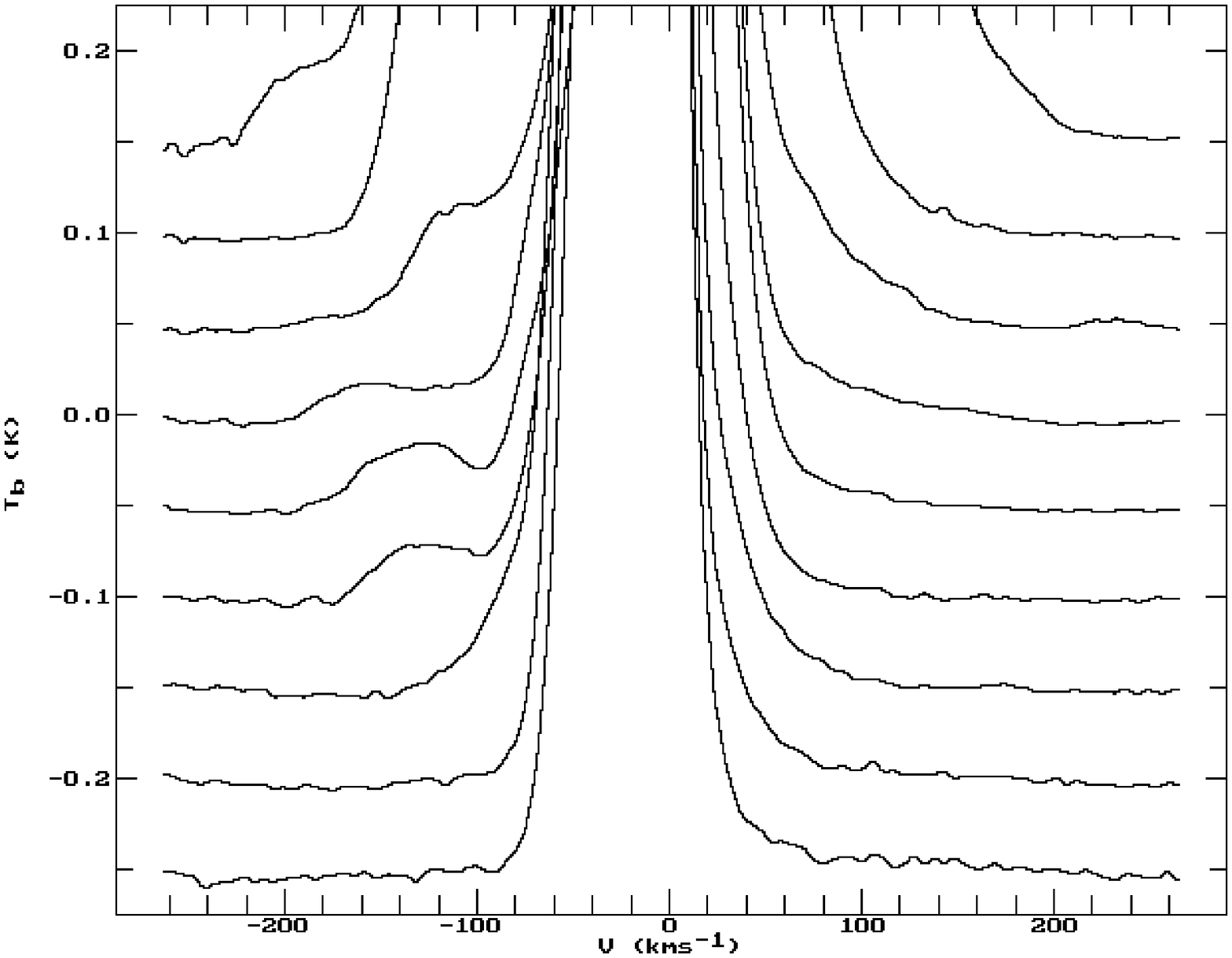,width=13.2cm,angle=-90
,bbllx=50pt,bblly=100pt,bburx=530pt,bbury=730pt}
\vspace*{-10.4875cm}
\centerline{\hspace*{-1.5cm}}
\psfig{figure=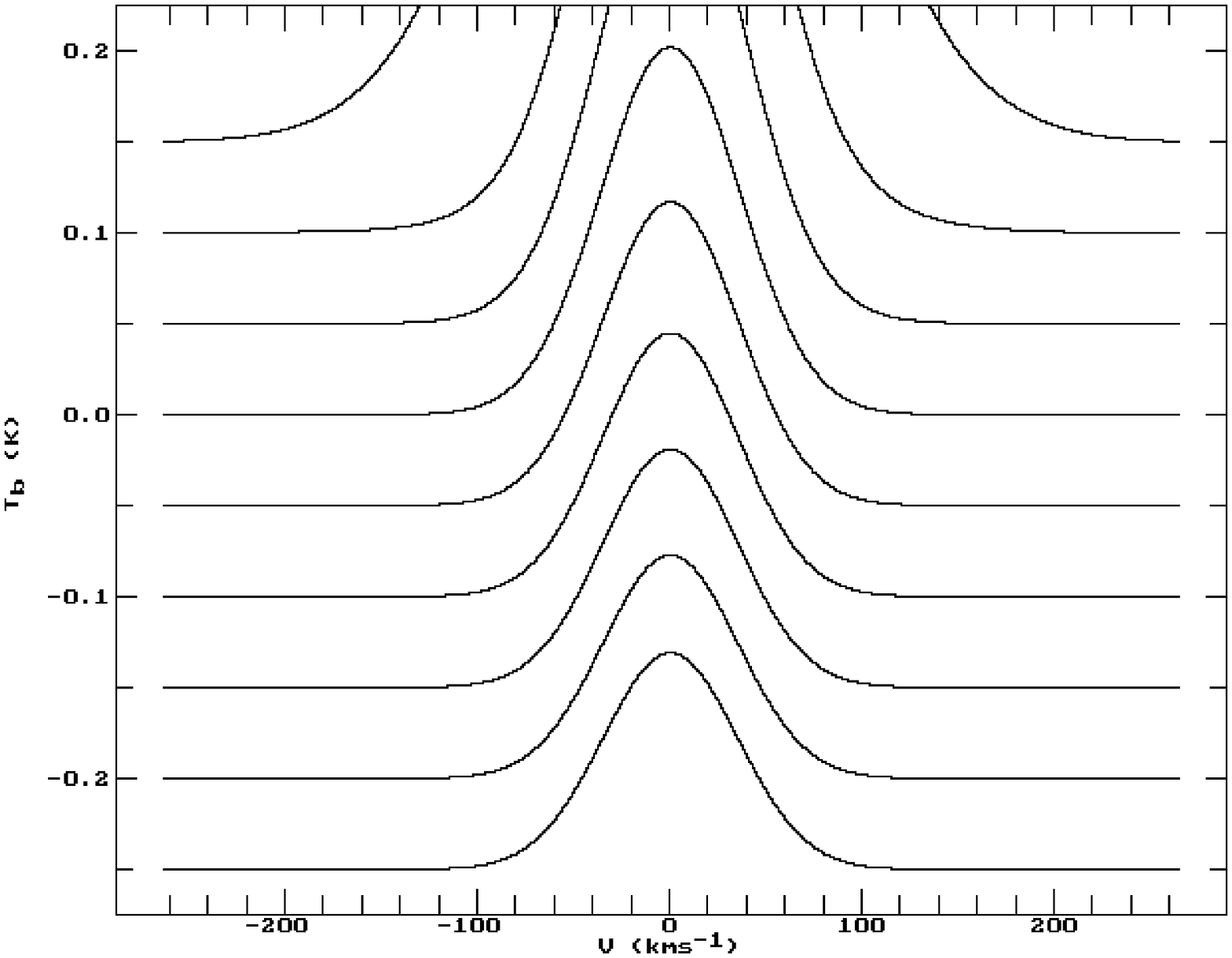,width=13.2cm,angle=-90
,bbllx=50pt,bblly=100pt,bburx=530pt,bbury=730pt}
\caption[]{
\hi 21-cm line spectra averaged across all longitudes as in Fig. 1, but extracted
from the Bell Labs Survey (BLS).
The solid lines represent the neutral atomic hydrogen layer model of Lockman \& Gehman (1991, 
their
model 2).
\label{fig2} }
\end{figure}
                                      
The resulting BLS baselines are believed to be accurate to 50 mK. 
Averaging observations in time should improve the baseline accuracy.
However, the averaging of \hi spectra of the BLS does not show this expected
behavior (Stark et al. 1992).
In  opposite to this finding, the individual spectra of the LDS are noisier
and have a typical rms noise level of 70 mK.
Averaging the LDS spectra however 
leads to the expected improvement of the rms noise up to a limit 
of a few mK.  

In comparing BLS and LDS spectra after averaging across regions of 
$5\deg\,\times\,5\deg$ Kalberla et al. (1998b) found significant 
deviations, indicating incompatible baselines between both surveys.

The majority of the averaged BLS \hi spectra reveal negative brightness 
temperatures in the baseline regions. Kalberla et al. (1998b) concluded, 
that probably an improper baseline correction algorithm was 
used by the reduction of the BLS data. In turn this 
led to a subtraction of the faint and broad \hi line emission of
the high-velocity dispersion component.
Polynomial baseline corrections are designed to {\em remove} features
from the \hi spectra which are assumed to be artificial.
Accidentally, broad emission line components may be affected.

The question, whether the baseline correction procedure applied to the 
LDS may have {\em added} an artificial broad emission line component 
was studied by Kalberla et al. (1998a) in detail. The complete LDS 
data analysis was repeated. No evidence was found that the high-velocity 
dispersion component is due to instrumental or computational artifacts. 
 
{\it Data analysis:\/} 
Both surveys have been analyzed regarding low surface brightness components with 
high-velocity dispersion. In case of the BLS Kulkarni \& Fich (1985) as well 
as Lockman \& Gehman (1991) averaged large regions in direction to the 
Galactic poles ($|b| > 80 \deg $ and $|b| > 45 \deg $). Such averages 
were decomposed into Gaussian components. 

Westphalen (1997) 
considered individual $10\deg\,\times\,10\deg$ fields in all directions with 
latitudes $|b| > 30 \deg $. After a Gaussian decomposition of the individual 
averaged profiles, the average 
property of a component with high-velocity dispersion was determined. 
In each case the rms deviations between the individual observations have 
been determined on a channel by channel basis; rms-peaks indicating 
interference, HVCs or instrumental problems have been used to flag 
velocity intervals, which need not to be fitted by a Gaussian component. 
To verify, that the determined components are not affected by residual stray 
radiation, the stray radiation itself was decomposed into Gaussian components. 
Since a Gaussian analysis may be biased by initial constraints, each fit 
was repeated several times using different criteria for the derivation of 
components with high-velocity dispersion. 


To compare the BLS and LDS datasets and the models 
based on a high-velocity dispersion component, we plot data and 
corresponding models. Fig. 1 is based on LDS data. We include the 
model derived by Kalberla et al. (1998a). Fig. 2 represents the BLS data
and the model 2 published by Lockman \& Gehman (1991) for comparison.
                
\section{Halo models}
The basic hydrostatic halo model is due to Parker (1966). In this 
classical publication it was shown that a single phase halo must be 
unstable. Since then, attempts have been made by various authors
to derive conditions for a stable halo. We consider only the most recent 
papers. 

Bloemen (1987) concluded that a gaseous halo is needed to 
provide the pressure necessary to stabilize the halo. He 
proposed an extended high temperature halo. At a vertical
distance of $1\,<\,|z|\,<\,3$\,kpc the halo gas temperature should be
$T\,\simeq \,(2-3)\; 10^5$\,K while within 
the disk and above $|z|\,>\,3$\,kpc the
temperature should be $T\,\simeq \,10^6$\,K.
This halo plasma should emit significant
amounts of X-ray photons within the 1/4\,keV and 3/4\,keV energy bands.

Boulares \& Cox (1990) studied the Galactic hydrostatic equilibrium
including magnetic tension and cosmic-ray diffusion.
They found, that the gas pressure, based on recent observations at that time,
was not sufficient to stabilize the halo against Parker instabilities.
For a stable Galactic halo, they proposed a halo \hi phase with a velocity 
dispersion of
20 \kmswss in the disk ($z\,=\,0$) and of 60 \kmsws at $z = 4 $ kpc.
This is close to the high-velocity dispersion component established 
by Kalberla et al. (1998a). 
The LDS observations indicate however, that the \hi emission in the halo
seems to be isothermal with a velocity dispersion which is 
independent from $z$-height.
We calculated the \hi distribution as proposed by Boulares \& Cox (1990) 
and found that this distribution is similar 
to the distribution derived from the Lockman \&  Gehman (1991) model which is 
plotted in Fig. 2. 

\subsection{A new approach}

To describe the large scale properties of the Milky Way in the sense of a
steady-state solution, Kalberla \& Kerp (1998)
assumed a hydrostatic halo model, as proposed by
Parker (1966).
Kalberla \& Kerp (1998) re-analyzed the hydrostatic equilibrium model 
of the Milky Way, including
the most recent all-sky surveys, ranging from the $\gamma$-ray to the 
radio synchrotron emission regime.
They distinguished three main components of the Galaxy which can 
be characterized by their
vertical scale heights $h_z$:
the cold and warm neutral interstellar medium (ISM) with 
$h_z$ = 150 pc and $h_z$ = 400 pc
(Dickey \& Lockman, 1990), the diffuse ionised gas (DIG) with $h_z$ = 1 kpc 
(Reynolds, 1997),
and halo gas with $h_z$ = 4.4 kpc.
The major difference to previous investigations (e.g. Bloemen 1987,
Boulares \& Cox 1990) is, that Kalberla \& Kerp (1998) included the neutral 
(Kalberla et al. 1998a)
and X-ray (Pietz et al. 1998) gas phase, located within the Galactic halo, into their 
calculations.
Such a layered disk-halo model was found to fit the large scale distribution
of the neutral and ionized hydrogen gas throughout the whole Galaxy.
Gas, magnetic fields and cosmic rays were found to be in pressure equilibrium.
                                                                         
As discussed by Kalberla \& Kerp (1998) in detail, such an equilibrium
situation does not only explain the large scale distributions of gas,
magnetic fields and cosmic rays, but results also in a stable configuration.
It is essential, that the multi-phase halo is supported by the pressure of the
gaseous layers with low $z$ scale-heights, belonging to the Galactic disk
and to the disk-halo interface. Without such layers, the halo would
be instable as described by Parker (1966).
                                          
Kalberla \& Kerp (1998) found that the halo is stable on {\em average} only.
Pressure fluctuations may cause instabilities above $z$ distances of
$\sim$ 4 kpc, but not below. Considering activities in the Galactic
disk (e.g. Norman \& Ikeuchi 1989) one expects that such pressure fluctuations
will affect the halo gas. Necessarily, instabilities will trigger the
condensation of HVCs in the halo. Kalberla et al. (these proceedings) 
studied the large scale distribution of HVCs and found that turbulent 
motions within the halo have considerable effects on the observed 
column density distribution. Except from turbulence, 
the scenario, derived from hydrostatic considerations,
is in close agreement with predictions of hydro-dynamical models.
Fountain parameters which have been favored by Bregman (1980)
are in close agreement with the parameters derived by Pietz et al. (1998)
from observations. 
                  
A gaseous multi-layer structure was found also by Avillez (these proceedings) 
from 3D hydro-dynamical simulations. In his calculations a halo, and at the
same time a disk-halo interface is built up after a short period of time,
reaching a steady state situation soon.
Scale heights, densities and temperatures of these layers are found to
be comparable to the values listed by Kalberla \& Kerp (1998).
                                                              

\end{document}